\PassOptionsToPackage{usenames, dvipsnames, table, xcdraw}{xcolor}
\documentclass[sigplan, screen]{acmart}
\renewcommand\footnotetextcopyrightpermission[1]{}

\usepackage{mathptmx} %

\usepackage{fancyhdr}
\usepackage[normalem]{ulem}

\newcommand{\microsubmissionnumber}{757}

\fancypagestyle{firstpage}{
  \fancyhf{}
  
  \fancyhead[C]{\vspace{15pt}\normalsize{MICRO 2020 Submission
      \textbf{\#\microsubmissionnumber} -- Confidential Draft -- Do NOT Distribute!!}} 
  \fancyfoot[C]{\thepage}
}

\usepackage{textcomp}
\usepackage[table,xcdraw]{xcolor}
\usepackage{booktabs} %
\usepackage{listings} %
\usepackage[english]{babel} %
\usepackage{siunitx}

\usepackage{algorithm}
\usepackage{algpseudocode}
\usepackage{amsmath}
\usepackage{amsfonts}
\usepackage{array}
\usepackage{boxedminipage}
\usepackage{calrsfs}
\usepackage{caption}
\usepackage[capitalize]{cleveref}
\usepackage{color}
\usepackage{enumitem}
\usepackage{fancyhdr}
\usepackage{float}
\usepackage{graphicx}
\hypersetup{
    colorlinks=false,
    pdfborder={0 0 0},
}
\usepackage[utf8x]{inputenc}
\usepackage{multirow}
\usepackage{subcaption}
\usepackage{verbatim}
\usepackage{xspace}
\usepackage{tikz} %

\newif\ifshowcomment
\showcommenttrue

\ifshowcomment

\newcommand{\todo}[1]{\noindent\textsf{\color{orange}{[{{\bf \scalebox{0.75}{\fbox{ToDo}}}: {\it#1}]}}}}
\newcommand{\dan}[1]{\noindent\textsf{\color{Violet}{[{{\bf \scalebox{0.75}{\fbox{Dan}}}: {\it#1}]}}}}
\newcommand{\vijay}[1]{\noindent\textsf{\color{purple}{[{{\bf \scalebox{0.75}{\fbox{Vijay}}}: {\it#1}]}}}}
\newcommand{\vasilis}[1]{\noindent\textsf{\color{orange}{[{{\bf \scalebox{0.75}{\fbox{Vasilis}}}: {\it#1}]}}}}
\newcommand{\arpit}[1]{\noindent\textsf{\color{magenta}{[{{\bf \scalebox{0.75}{\fbox{Arpit}}}: {\it#1}]}}}}
\newcommand{\antonis}[1]{\noindent\textsf{\color{blue}{[{{\bf \scalebox{0.75}{\fbox{Antonis}}}: {\it#1}]}}}}

\else

\newcommand{\todo}[1]{}
\newcommand{\antonis}[1]{}
\newcommand{\dan}[1]{}
\newcommand{\vijay}[1]{}
\newcommand{\arpit}[1]{}
\newcommand{\vasilis}[1]{}
\fi

\newcommand{\beginbsec}[1]{\noindent\textbf{#1.}}

\newcommand{\squishlist}{
 \begin{list}{$\bullet$}
  { \setlength{\itemsep}{2pt}
     \setlength{\parsep}{0pt}
     \setlength{\topsep}{2pt}
     \setlength{\partopsep}{0pt}
     \setlength{\leftmargin}{1em}
     \setlength{\labelwidth}{1em}
     \setlength{\labelsep}{0.5em} } 
}

\newcommand{\squishlistContrib}{ %
 \begin{list}{$\bullet$}
  { \setlength{\itemsep}{2pt}
     \setlength{\parsep}{0pt}
     \setlength{\topsep}{2pt}
     \setlength{\partopsep}{0pt}
     \setlength{\leftmargin}{1em}
     \setlength{\labelwidth}{1em}
     \setlength{\labelsep}{0.5em} }
}
\newcommand{\squishend}{ \end{list}  }

\newcommand{\squishenum}{\begin{enumerate}[itemsep=0.5pt,parsep=0pt,topsep=0pt,partopsep=0pt,leftmargin=1.5em,labelwidth=1em,labelsep=0.5em]{}}
\newcommand{\squishenumend}{\end{enumerate}}

\newcommand\myurl[2]{\url{#1}}

\newcommand{\captionfonts}{\small}
\makeatletter  %
\long\def\@makecaption#1#2{%
  \vskip 0.1in
  \sbox\@tempboxa{{\captionfonts #1: #2}}%
  \ifdim \wd\@tempboxa >\hsize
    {\captionfonts #1: #2\par}
  \else
    \hbox to\hsize{\hfil\box\@tempboxa\hfil}%
  \fi
  \vskip 0in}
\makeatother   %

\newcommand{\qt}[1]{``#1''}

\def\custvspace{\vspace{0.4em}}

\def\kv{\emph{KV-pair}}
\def\kvs{\emph{KV-pairs}}
\def\state{\emph{state}}

\def\acclogno{\emph{accepted-log-no}}
\def\logno{\emph{log-no}}
\def\lognos{\emph{log-nos}}
\def\lognoeq{\emph{log-no =}~}
\def\lognox{\emph{log-no = X}}
\def\lognoxmin{\emph{log-no = X - 1}}

\def\comlogno{\emph{last-committed-log-no}}

\def\comlognoxmin{\emph{last-committed-log-no = X - 1}}

\def\val{\emph{value}}
\def\accval{\emph{accepted-value}}
\def\helpflag{\emph{helping-flag}}
\def\helploc{\emph{helping-local-entry}}
\def\allab{\emph{all-aboard}}
\def\allabtimeout{\emph{all-aboard-time-out-counter}}

\def\propts{\emph{proposed-TS}}
\def\accts{\emph{accepted-TS}}

\def\RMW{RMW}
\def\RMWs{RMWs}
\def\rmw{\emph{rmw-id}}
\def\rmws{\emph{rmw-ids}}
\def\comrmw{\emph{last-committed-rmw-id}}

\def\loghigh{\emph{Log-too-high}}
\def\loglow{\emph{Log-too-low}}
\def\alreadycom{\emph{Rmw-id-committed}}
\def\highprop{\emph{Seen-higher-prop}}
\def\highacc{\emph{Seen-higher-acc}}
\def\lowacc{\emph{Seen-lower-acc}}
\def\ack{\emph{Ack}}
\def\acks{\emph{Acks}}
\def\ackbase{\emph{Ack-base-TS-stale}}

\def\locentry{\emph{Local-entry}}
\def\locentries{\emph{Local-entries}}
\def\retry{\emph{Retry-with-higher-TS}}
\def\need{\emph{Needs-KV-pair}}
\def\bcast{\emph{Bcast-commits}}
\def\bcasthelp{\emph{Bcast-commits-from-help}}
\def\committed{\emph{Committed}}

\def\regedrmw{\emph{last-registered-rmw-id}}

\def\basets{\emph{base-TS}}
\def\accbasets{\emph{acc-base-TS}}

\def\invalid{\emph{Invalid}}
\def\proped{\emph{Proposed}}
\def\acced{\emph{Accepted}}

\def\carstamplow{\emph{Carstamp-too-low}}
\def\carstampequal{\emph{Carstamp-equal}}
\def\carstamphigh{\emph{Carstamp-too-high}}

\newcommand{\secref}[1]{Section~\ref{#1}}

\def\eg{e.g.,~} %
\def\ie{i.e.,~} %

\title[Extending Classic Paxos for High-performance RMW Registers]{Extending Classic Paxos for High-performance Read-Modify-Write Registers}

\author{Vasilis Gavrielatos, Antonios Katsarakis, Vijay Nagarajan}
\affiliation{%
  \institution{The University of Edinburgh\\
  FirstName.LastName@ed.ac.uk}
}

\begin{document}

\maketitle
\pagestyle{plain}

\section{Introduction}

In this paper we pose the problem of executing Read-Modify-Writes (\RMWs) over replicated data, assuming that machines can crash and that processing and networking delays can be unbounded.
This is also known as solving \emph{asynchronous consensus}.
Only that instead of assuming that machines must reach agreement on the next value of a \qt{register}, we assume that machines agree on the next \RMW\ to be executed over a replicated key-value pair that resides within an in-memory Key-Value Store (KVS).
We also pose two additional requirements.

Firstly, we are looking for a solution that does not sacrifice availability when a machine fails. This renders the majority of consensus protocols unsuitable for our purpose. For instance, consider leader-based~\cite{Lamport:2001, Ongaro:2014, Reed:2008} consensus protocol. When the leader fails the rest of the machines must block waiting for a timeout to expire before performing leader election. Such protocols sacrifice availability for the duration of that timeout.

Secondly, we want to deploy the system in the datacenter over modern hardware. This means that the protocol must scale across the many cores of a modern server and must be able to utilize the high bandwidth of modern network interfaces. This requirement rules out the last standing protocol, EPaxos~\cite{Moraru:2013}, that can solve asynchronous consensus without availability penalties. This is because EPaxos is so complex that is not clear how it can be deployed in a scalable fashion across the many threads of each machine. Absent that thread-scalability, it is impossible to leverage the high bandwidth offered by modern networks.

\custvspace
The requirements of 1) solving asynchronous consensus, 2) avoiding any availability penalty and 3) leveraging modern hardware,  all point to 
Classic Paxos~\cite{Lamport:1998} (from now on \emph{CP} or \emph{Paxos}), which solves asynchronous consensus without blocking on a machine crash or delay while it can be deployed on a per-key basis without requiring any synchronization across threads. However there is a problem.
For each key, CP can decide only a single value once. 
However, we must be able to perform many \RMWs\ on the same key.
Therefore, to use CP, we must first extend it, such that it can run repeatedly for each key, while preserving important invariants such that each \RMW\ executes exactly once. Finally, we must make sure to avoid livelocks, which are a known issue of CP.

\custvspace
In this work we provide a detailed specification of how we extended and implemented Classic Paxos to execute Read-Modify-Writes in Kite~\cite{V:2020}. In addition, we also specify how we implemented All-aboard Paxos~\cite{Howard:2019} over Paxos and how we use carstamps~\cite{Burke:2020}, to also add ABD reads and writes, to accelerate the common case, where \RMWs\ are not needed.
Our specification targets a Key-Value-Store that is deployed within the datacenter, is replicated across 3 to 7 machines and supports reads, writes and \RMWs.

First, to allow CP to run repeatedly we use the abstraction of a log per key, but -- crucially -- without actually implementing a log per key. Specifically, every time we commit an \RMW\ for a key, we conceptually move to the next log entry, and we denote so through a single counter that remembers the current entry. The log abstraction allows us to execute CP repeatedly across the different entries, without actually modifying CP. The modularity of this approach is crucial as it allows us to rely on the well-established correctness guarantees of CP.

Second, once we have established the log abstraction, we must ensure that an \RMW\ that has been committed in log entry $X$ can never be committed again in log entry $Y$, $Y > X$. We achieve this by tagging \RMWs\ with unique identifiers (\rmws), and remembering the latest \rmw\ committed from each session (bounded storage). In addition, we add commit messages to ensure that each replica knows when an \RMW\ has been committed, and can thus stop it from getting proposed again in the future.

Finally, recall that our goal is to create a multi-threaded system that can stress the  many-core servers and high-bandwidth NICs found in today's datacenter. To achieve this we need to allow for thousands of \RMWs\ to execute concurrently. However such concurrency can be problematic with CP as it can trigger livelocks. We avoid this scenario by implementing a back-off mechanism that for each key, allows each machine to have at most one active \RMW. Furthermore, if machine M1 sees that machine M2 is trying to perform an \RMW, then M1 will back-off, giving M2 a chance to complete its RMW unencumbered.

\custvspace
Beyond extending CP, we also provide the specification of an optimization called All-Aboard Paxos, which we implemented over our CP implementation. All-aboard is an optimization sketched in Howard's thesis~\cite{Howard:2019} as an application of the Flexible Paxos~\cite{Howard:2016} theorem. To the best of our knowledge we are the first to fully specify and measure an All-aboard implementation.
Finally, to provide a system that offers a read, write, RMW API, we implement ABD reads and writes using carstamps~\cite{Burke:2020} over our CP implementation.

The rest of this paper is organized as follows.
\secref{sec:cp} briefly describes how CP works (without our extensions).
\secref{sec:prel} describes our execution model and the basic data structures that underpin our implementation.
\secref{sec:prot} specifies our implementation of CP following the lifetime of a request. \secref{sec:backoff} and \secref{sec:help} explain how back-off and helping are implemented. \secref{sec:cor} provides a series of correctness arguments and informal proofs for our extensions. \secref{sec:why} provides the rationale behind an array of our design choices. \secref{sec:all-aboard} presents the theory and implementation of All-aboard Paxos. \secref{sec:writes} discusses how we added ABD writes to our implementation using carstamps~\cite{Burke:2020}. Finally, \secref{sec:reads} describes how we added ABD reads.

\section{Classic Paxos Algorithm} \label{sec:cp}

Here we will give a very brief overview of Classic Paxos through Table~\ref{tab:CP}.
We will not explain why it works, but rather just how it works.
There are two types of messages: proposes and accepts. A machine first issues proposes, if it gathers a majority of acks it then issues accepts. If accepts are acked by a majority of machines then the command is said to have been decided (\ie committed).

Proposes and accepts carry logical timestamps (\ie TSes, discussed in~\ref{sec:prel:data}).
Table~\ref{tab:CP} shows the case where machine M1 sends messages and M2 receives them.
The leftmost column shows four potential messages that M1 can send, and the top row shows the four possible states that M2 can be in. We assume two timestamps \emph{H} and \emph{L}, where H > L. 
The rest of the cells denote the answer that M2 sends to M1 based on  M1's message and M2's state. 
Each cell is colour-coded as blue, red or green, allowing us to group similar cases. Below we discuss each of the colours, noting that the crux of the protocol is found in the red cells.

\custvspace
\beginbsec{Green cells}
The green cells are when an accept message of M1 finds M2 having only received a propose with the same TS. This is the most straight-forward case where there has not been another intervening command, and M2 will always accept and reply with an ack.

\custvspace
\beginbsec{Blue cells} Blue cells capture obscure corner cases, such as propose-L finds that M2 has already seen a propose-L. This is obscure because TS should be unique and we should expect that no TS is used twice for proposes. These can occur for example when we send the message twice, because we suspect the first message was lost.
These cells are not of particular interest to understanding CP.

\custvspace
\beginbsec{Red cells} Finally, red cells shows how CP handles Conflicts between commands with different Timestamps. The crux of this protocol lays here.
Note the (simplified) rules.

\begin{enumerate}
\item Propose-H blocks propose-L and accept-L. I.e. proposes block proposes and accepts with lower TSes
\item Accept-H blocks propose-L and accept-L. I.e. accepts block proposes and accepts with lower TSes
\item Propose-L cannot block accept-H or Propose-H. I.e. proposes with lower TS are discarded
\item Accept-L cannot block accept-H, but it does block propose-H forcing it to help it. I.e. accepts with lower TSes cannot block accepts, but they do block proposes, and force them to help them
\end{enumerate}

Note that our rules assume that proposes and accepts will all be seen by a majority of nodes and thus will intersect. Later we will relax this assumption (see \S\ref{sec:all-aboard}).

\def\redcell{\cellcolor[HTML]{f4bfb4 }}
\def\greencell{\cellcolor[HTML]{ c1f4b4 }}
\def\bluecell{\cellcolor[HTML]{b4bbf4}}
\begin{table}[t]
\centering
\resizebox{0.48\textwidth}{!}{%
\begin{tabular}{|c|c|c|c|c|}
\hline
 &
\cellcolor[HTML]{C0C0C0}\begin{tabular}[c]{@{}c@{}}\textbf{M2}:  \\ Already seen \\ propose-L\end{tabular} &
\cellcolor[HTML]{C0C0C0}\begin{tabular}[c]{@{}c@{}}\textbf{M2}:  \\ Already seen\\ accept-L\end{tabular} &
\cellcolor[HTML]{C0C0C0}\begin{tabular}[c]{@{}c@{}}\textbf{M2}:  \\ Already seen\\ propose-H\end{tabular} &
\cellcolor[HTML]{C0C0C0}
\begin{tabular}[c]{@{}c@{}}
\textbf{M2}:  \\ Already seen\\ accept-H
\end{tabular} \\ 
\hline

\cellcolor[HTML]{C0C0C0}\begin{tabular}[c]{@{}c@{}}\textbf{M1}:  \\ 
Sends propose-L 
\end{tabular} &
\bluecell\begin{tabular}[c]{@{}c@{}}Nack-\\ restart\end{tabular}& 
\bluecell\begin{tabular}[c]{@{}c@{}}Nack-\\ restart\end{tabular}&
\redcell\begin{tabular}[c]{@{}c@{}}Nack-\\ restart\end{tabular}&
\redcell\begin{tabular}[c]{@{}c@{}}Nack-\\ restart\end{tabular} \\ 
\hline
\cellcolor[HTML]{C0C0C0}\begin{tabular}[c]{@{}c@{}}
\textbf{M1}:  \\ 
Sends accept-L 
\end{tabular}&
\greencell Ack &
\bluecell Ack&
\redcell \begin{tabular}[c]{@{}c@{}}Nack-\\ restart\end{tabular} &
\redcell \begin{tabular}[c]{@{}c@{}}Nack-\\ restart\end{tabular} \\
\hline
\cellcolor[HTML]{C0C0C0}\begin{tabular}[c]{@{}c@{}}\textbf{M1}:  
\\ Sends propose-H 
\end{tabular}&
\redcell  Ack &
\redcell\begin{tabular}[c]{@{}c@{}}Nack -\\ Help\end{tabular}&
\bluecell\begin{tabular}[c]{@{}c@{}}Nack-\\ restart\end{tabular}&
\bluecell\begin{tabular}[c]{@{}c@{}}Nack-\\ restart\end{tabular}\\ 
\hline
\cellcolor[HTML]{C0C0C0}\begin{tabular}[c]{@{}c@{}}\textbf{M1}:  \\ Sends accept-H \end{tabular}&
\redcell Ack  &
\redcell Ack  &
\greencell Ack  &
\bluecell Ack\\ 
\hline
\end{tabular}
}
\caption{A simplified version of the core CP algorithm}
\label{tab:CP}
\end{table}

\section{System model and data structures} \label{sec:prel}

There are a finite number of machines (aka servers or nodes). Typical numbers are 3 to 7. Each machine runs a number of worker software threads (\emph{workers}). Typical numbers are 20 to 30. Each worker runs a number of sessions. Typically 40 to 80. Each session maps to an external client. There is one FIFO per session. Client requests are inserted in the FIFO.
Workers execute the requests of each session in order. Requests from different sessions can run concurrently. We will assume that all requests are read-modify-writes (RMWs). 

Therefore, at any given moment the worker can be working concurrently on one RMW for each session.
With 20 workers, 40 sessions and 5 machines, at any given moment there are 4000 RMWs running concurrently.

Each machine has the same Key-Value Store (KVS) in its main memory.
For each RMW a Paxos instance is running. RMWs to different \kvs\ do not have any relation whatsoever. RMWs to the same \kv\ are said to \emph{conflict}.

\subsection{Data-structures and preliminaries}~\label{sec:prel:data}
We start by establishing the data structures that will be used by the implementation.

\custvspace
\beginbsec{Message Types}
Paxos requires 2 broadcast rounds, a \emph{propose} broadcast followed by an \emph{accept} broadcast. Finally, if the accept is successful, a \emph{commit} message will be broadcast.

\subsubsection{The \kv~and its metadata}
\custvspace
\beginbsec{\kv}
For each key in the KVS, the KVS stores a data structure that contains the key, the value and some metadata. We will refer to this structure as \emph{\kv}. The \kv~is where RMWs to the same key from different machines/workers/sessions meet and synchronize. The metadata of the \kv\ will describe the state of the executing RMW (if any) at any given moment as perceived by the machine that stores the \kv.

\custvspace
\beginbsec{\kv~fields}
Below is a list of all the fields of \kv. The rest of this subsection is devoted to explaining these fields.
\squishenum
\item key
\item value
\item accepted-value
\item state
\item log-no
\item last-committed-log-no
\item proposed-TS
\item accepted-TS
\item rmw-id
\item last-committed-rmw-id
\squishenumend

\custvspace
Crucially, every piece of metadata is absolutely necessary as we will see throughout this paper.

\custvspace
\beginbsec{State}
Each \kv\ stores a \state\ variable. The state can be \invalid, \proped\ or \acced.
The \state\ variable refers to the last not-yet-committed \logno. 
\invalid\ means that the \kv~knows of no ongoing RMW attempt. On receiving a propose (locally or from a different machine) the state will go to \proped. On receiving an accept the state will go to \acced. As we explain the protocol we will demonstrate when such state transitions are legal.

\custvspace
\beginbsec{Logical Timestamps---\emph{TSes}}
Paxos makes heavy use of logical timestamps (aka Lamport Clocks). A timestamp is a tuple of a version and a machine-id. To compare two timestamps, we compare their versions, using the machine-ids as tie breakers. We will refer to logical timestamps simply as \emph{TSes}.

\custvspace
\beginbsec{\kv~Timestamps}
As part of the \kv's metadata there are two TSes. A \propts, which remembers the highest propose seen and an \accts, which remembers the highest TS that has been accepted. The \accts\ is used only in one case: when the \kv~is in \acced\ state and a propose with a lower TS than the \accts\ is received.

\custvspace
\beginbsec{Log number - \logno}
One of \kv's metadata fields is the \logno, which is a counter for the number of \RMWs\ that have committed for that particular \kv. Despite its name, there is no actual log -- or need for one -- in the implementation~\footnote{Interestingly Lamport also observes that there is no need to keep an actual log for a KVS in his original paper~\cite{Lamport:1998}.}. However, it is useful to think of a log for each \kv; for each of the log's slots, Paxos must run to decide the winner that gets to commits its \RMW. 

Therefore, when a \kv\ stores a \lognoeq 10, that means that the key has already been successfully RMWed 9 times (at least), and we are currently working on the tenth. We do not actually keep a log, because we need not remember any of the values committed in slots 1 - 8.  
The fields: state, \propts, \accts, \accval\ and \rmw, all refer to the \logno, \ie to the log number we are currently working on. If the state is \invalid, these fields are meaningless. 
The \kv\ also stores a \comlogno, that refers to the most recent \logno\ that has been committed (that it knows of). The \val\ always refers to the value committed in \comlogno.

\custvspace
\beginbsec{Helping}
In Paxos it is possible for machine M1 to \qt{help} an \RMW\ of a different machine M2. 
The help is necessitated by the uncertainty of the asynchronous environment: M1 cannot always know whether M2 has failed or not, and it cannot always know if an \RMW\ is committed or not (because the \RMW\ may have been committed by a majority that M1 cannot reach in its entirety).

\custvspace
\beginbsec{RMW-ids}
As a result of helping, it is possible that machine M2 attempts to commit its \RMW, even though machine M1 helped that \RMW\ in the past. Therefore, if we are not careful it may be the case that M1 commits an \RMW\ in \lognoeq 10 and then machine M2 completes the exact same \RMW\ in \lognoeq 11.
This violates correctness. To ensure that an \RMW\ is committed exactly once, we use \rmws. An \rmw\ is an 8-Byte value, the LSBs of which contain the global-session-id (it could also be a tuple of <id, global-session-id>). Each session in the system has its own global-session-id and thus each \RMW\ is granted a unique \rmw. This allows us to identify different \RMWs.

\custvspace
\beginbsec{Registering \rmw}
Each machine keeps the latest \RMW\ that it knows has been committed by each session globally.
Therefore, with 5 machines, 20 workers, 40 sessions, each machine holds an array with 800 fields. Each field denotes the most recent \rmw\ of every other session. Note that if we know that the \rmw\ <10, 253>, from session 253, has been committed, then it must be that the \RMW\ with \rmw\ <9, 253> has also been committed. Therefore, we need only remember the latest \rmw\ committed by each session. When a worker learns about a committed-id, it immediately \emph{registers} it, i.e. it updates the array with committed \RMWs\ accordingly.

As a result when a machine tries to commit an already committed \RMW, other machines are able to detect it and reply to it that the \RMW\ has already been committed.

\custvspace
\beginbsec{Necessary \rmws\ in the \kv}
The \kv~stores: 1) an \rmw, referring to the the \RMW\ currently being worked on the \logno\ and 2) a \comrmw~which refers to the last committed RMW  (in \comlogno).

\custvspace
\beginbsec{Accepted-value}
In order to facilitate help, we must know the result of the \RMW\ we are helping, this is why the \kv\ has the field \accval, to remember the value that the highest accepted \RMW\ wants to commit.
The \accval\ field is only valid, if the key is in \acced\ state and it only refers to the \logno. Note that alternatively, we could only be storing the identity of the \RMW\ operation. For example, if the \RMW\ is a Fetch-and-Increment, we can store its opcode instead of the value it creates, and calculate the value on demand (although the value is probably no more than 8 bytes in this scenario). 
However, this implementation assumes that the most commonly \RMWs\ are Compare-and-Swaps, where remembering the whole identity of the \RMW\ would be twice as expensive as storing its result (it has a compare-value and an exchange-value!).

\subsubsection{Local-entries} \label{sec:base:loc}
\custvspace
\beginbsec{Local-entries} Each worker has one \locentry\ pre-allocated for each session.
The \locentry\ holds all the necessary state for the RMW. It remembers the exact operation the RMW wants to achieve and to which key, whether the RMW has communicated with the local \kv\ or other machines, what types of responses it has received and so on. Note the contrast between a \kv\ and a \locentry: the \kv\ is shared among all worker threads and stores the state of the RMW that is currently in the front stage executing; \locentries\ are thread-local and store the state of every RMW, including that of RMWs that are currently sidelined, in a back-off state waiting to get access to the \kv.

The \locentry\ has a state variable, describing the state of the RMW: the state can be \invalid, \proped, \acced, \need, \retry, \bcast, \bcasthelp\ and \committed. We will discuss these states as we go along. Here is a list of field members of the \locentry.
\squishenum
\item \emph{key}
\item \emph{state}
\item \accval
\item \acclogno
\item \rmw
\item \emph{back-off-counter}
\item \helpflag
\item \helploc
\squishenumend

\custvspace
\beginbsec{Unique Identifier -- \emph{lid}}
When receiving a reply to a message we need to know to which RMW it corresponds. But because the same RMW may have triggered multiple broadcasts of the same type, we tag each broadcast with a unique identifier called \emph{lid}. The \locentry~remembers the lid used for the most recent broadcast (propose or accept). 
Replies to that broadcast always include the lid. On receiving a reply, the worker searches if there is an active \locentry~with that lid. If not it disregards the message. Otherwise, the worker passes the reply to the \locentry~. As an optimization, we use the least significant bits of the lid, to store the session-id, such that we can immediately locate the  \locentry, a reply refers to, without the need for a (linear) search.

This is not merely a networking implementation detail. A unique identifier is necessary because it allows us to discard replies to older messages, \eg a reply for the same session, same \rmw~and \logno, may refer to an older propose attempt. 

\custvspace
\beginbsec{Jargon} We will use the term local/locally to refer to the issuing machine of an RMW, or the machine that stores the KVS. For example, an RMW is said to be locally accepted, if it has gotten the \kv~that is stored in the issuing machine's KVS in \acced\ state.

\subsubsection{Execution Model}

The worker operates in a while(true) loop. In every iteration it: 1) polls for remote messages and takes action based on them by altering local state and enqueuing replies, 2) inspects all active \locentries~(i.e. not in \invalid~state), and it takes action based on their state, 3) it sends enqueued messages (either broadcasts or unicasts which are always replies to broadcasts) and 4) probes the client FIFOs to pull requests for sessions that are not blocked.
Therefore Paxos protocol actions are triggered, only in response to a received message or when inspecting the \locentry~of an RMW.

\section{The protocol spec}\label{sec:prot}
Here we will see the lifetime of an RMW step-by-step. The description includes a number of forward pointers explaining additional cases.

\subsection{Grabbing the local KV-pair}
When the worker pulls a new RMW for a session, it immediately takes it to the local KVS. If the \kv~is in \invalid~state, then the \kv~is \qt{grabbed} by the RMW, transitioning it in \proped~state and essentially stopping other RMWs from the same machine, from attempting RMWs. 
If the \kv~is already grabbed, then the RMW must \emph{back-off} by transitioning its \locentry~to a state called \need.

Note that if the \kv~can be in \invalid~state then the RMW on the previous \logno~has been committed:
\kv.\comlogno~+ 1 == \kv.\logno. (It is possible that an RMW has worked on the current \logno~before, but it gave up and reverted it to \invalid~state-- discussed in \S\ref{sec:why:alreadycom}).

Let us assume that the RMW successfully grabs the \kv. (Section~\ref{sec:backoff} discusses the \need~state as part of the back-off mechanism.)
The RMW fills its corresponding \locentry, with information about its current state: the key to be RMWed, its \rmw, the \logno~it works on, and the TS it will use for the propose message. The TS.version can be any arbitrary number. We use the number 3 (this will become important when discussing the All-aboard optimization in Section~\ref{sec:all-aboard}). The TS.machine-id is the machine-id of the proposer.

\custvspace
\beginbsec{Propose}
The RMW must broadcast propose messages to the rest of the machines. A propose message includes the key, the TS of the propose, the \logno~and the \rmw. 

\subsection{On receiving a propose}
The receiver will go to its local KVS to examine the \kv~and sent back one of the following replies: \alreadycom, \loglow, \loghigh, \highprop, \highacc, \lowacc, or \ack.
Note that all replies that are not an \ack~are essentially a nack.
The replies always include an opcode.
Below we discuss the following: when each reply is triggered, what is its additional payload and what happens to the \kv~in the machine that sends the reply. 

\squishlist
\item \alreadycom: if the \rmw~of the propose message has been registered.  The reply message does not include additional payload. The \kv~is not altered. (Section~\ref{sec:why:alreadycom} elaborates on this.)

\item \loglow: if the \logno~of the propose message is smaller than the current working
\logno~of the \kv. This means the proposer does not know of the latest committed RMW, and thus attempts to commit on an already used \logno. The reply message includes the last committed RMW (\comlogno, \comrmw~and value). The \kv~is not altered.

\item \loghigh: if the \logno~of the propose message is higher than the current working \logno~of the \kv. This means the proposer knows of a committed RMW, that the receiver does not know of. The reply message does not include additional payload. The \kv~is not altered.

\item \highprop:  if the \kv~is in \proped~state but with a higher or equal \propts~than the propose's TS. The reply message includes the \propts~of the \kv. The \kv~is not altered.

\item \highacc: if the \kv~is in \acced~state but with a higher or equal \propts~than the propose's TS. The reply message includes the \propts~of the \kv. The \kv~is not altered. (This is identical as above.) (Note: the \accts\ is not inspected at all, because if the \propts\ is higher than the propose's TS, the proposer has to retry either way.)

\item \lowacc: if the \kv~is in \acced~state and its \accts~is lower than the propose's TS. Crucially, the \kv~remains in \acced~state, but its \propts~is advanced to the propose's TS, if it is smaller. The reply message includes the details of the accepted rmw, \ie  the \accts, the \rmw, and the \accval. 

\item \ack: if the \kv~is currently in \invalid~state, or the \kv~is in \proped~state with a lower \propts~than the propose's TS.
The \kv~goes to \proped~state and updates its proposed-TS. The reply message does not include any payload.
\squishend

\subsection{On receiving a propose-reply}
The replies are steered to local entries (as explained in \S\ref{sec:base:loc}).
The worker periodically polls active \locentries\ and takes action when the \locentry~signals that a majority (or more) of replies has been gathered or  one reply of type \alreadycom/\loglow/\highprop/\highacc\ has been received. With 5 machines we wait for 2 replies, since we already have a reply --typically an \ack, but not always-- from the local machine. Then the worker acts on the replies as follows.

Note that when processing the replies, there is a question of priority; \eg if there is 1 seen-higher-acc, 1 seen-lower-acc and 2 acks, what should we do? The following discussion takes that into account.

\squishlist

\item \alreadycom: if any \alreadycom~is received, then the proposer commits its RMW in the local \kv~(using the \locentry~fields \accval~and \acclogno). The \locentry~transitions to \bcast~state so that commits are broadcast (again using the \locentry~fields \accval~and \acclogno). \secref{sec:cor:inv3-nec} explains why using \accval\ is correct. \secref{sec:why:alreadycom} shows an optimization that allows us to avoid broadcast commits when not necessary.

\item \loglow: if any \loglow~is received, the RMW it contains is committed in the local \kv.
If none of the above replies have been received, then the \locentry~transitions to \need~state, so that it tries to grab the entry again, but in a later \logno. Section~\ref{sec:why:loglow} elaborates on this.

\item \highprop/\highacc: if a \highprop\ or \highacc~is received and none of the above is received then the \locentry~transitions to \retry~state. The RMW then will retry to broadcast proposes using a TS that is higher than all TSes included in the received replies. Section~\ref{sec:how:retry} describes in detail the different cases when retrying.

\item \ack: if a majority of \ack~replies has been gathered and none of the above is received, then the RMW will attempt to accept locally, by transitioning the local \kv~to \acced~state.
Section~\ref{sec:how:accept} describes in detail the different cases when accepting locally.
If it is successful, it will then calculate the \accval, store it in the \locentry~and the \kv, store the \logno~in the \locentry~as \acclogno~and broadcast accepts. The accept messages will include, the \accval, the \logno, the \rmw~and the same TS as the propose.

\item \lowacc: If a \lowacc~reply has been gathered and none of the above conditions have been triggered, then we will attempt to help the received RMW. Section~\ref{sec:how:accept} describes in detail how helped RMWs are accepted locally. After accepting locally, accepts for the helped RMW are broadcast.

\item \loghigh: If a \loghigh~reply has been received and none of the above has occurred (\eg with 5 machines we have received 1 \loghigh~and the 2 \acks), then we retry the RMW, transitioning the \locentry~to \retry-state. Retrying is discussed at length in Section~\ref{sec:how:retry}. Notably, if this occurs repeatedly, we will timeout, broadcasting commits from the immediately previous \logno. This is discussed in detail in \secref{sec:loghigh} .

\squishend

\custvspace\beginbsec{Priority}
The priority is due to performance and not correctness.
If the RMW has been committed, we should not waste our time with other replies.
If we are working on an already committed \logno, the same.
If our TS is too small, we should not bother checking acks, we will probably end-up loosing anyway. Similarly, if acks have been gathered, we should not try to help a lower-TS accept, as we are sure to have blocked it. Finally, we should only care about the \loghigh~reply, only if none of the above is triggered.

\subsection{Accept}
After managing to accept locally, accept messages are broadcast to remote machines.
Note that when accepting locally the value-to-be-written and value-to-be-read for the RMW are calculated. The value-to-be-written is called \accval. The \accval, must be stored in both the \kv's metadata in the KVS and the \locentry. 
The \kv~needs the \accval~to facilitate help, by responding to proposes with higher TS with the \lowacc~reply, that includes the \accval. The \locentry~needs the \accval, both to create accept messages that will be broadcast, but also in case it gets helped by another machine, to know what value must be read. Help is discussed in Section~\ref{sec:help}, reading the correct value is discussed in Section~\ref{sec:cor:inv3-nec}.
The accept messages will include, the \accval, the \logno, the \rmw~and the same TS as the propose.

\subsection{On receiving an accept} \label{sec:prot:acc-recv}
The receiver will go to its local KVS to examine the \kv~and send back one of the following replies:
\alreadycom, \loglow, \loghigh, \highprop, \highacc~or \ack.
Note that all replies that are not an \ack~are essentially a nack.
The replies always include an opcode.
Here is when each reply is triggered, what it is its additional payload and what happens to the \kv~in the receiving side. A lot of the replies are identical to the ones described for the propose, in that case we will simply denote so.
\squishlist
\item \alreadycom: <identical to propose>
\item \loglow: <identical to propose>
\item \loghigh: <identical to propose>

\item \highprop:  if the \kv~is in \proped~state but with a higher (but not equal, this is the difference with propose-replies) \propts~than the accepts's TS. The reply message includes the \propts~of the \kv. The \kv~is not altered. 

\item \highacc: if the \kv~is in \acced~state but with a higher (but not equal, this is the difference with propose-replies) \propts~than the accept's TS. The reply message includes the \propts~of the \kv. The \kv~is not altered. (This is identical as above.)

\item \ack: if the \kv~is currently in \invalid~state, or the \kv~is in \proped~state with a lower or equal \propts~than the accept's TS, or the \kv~is in \acced~state with a lower or equal \propts~than the accept's TS.
The \kv~goes to \acced~state and updates its \propts, \accts~and \accval. The reply message does not include any payload.
\squishend

\subsection{On receiving an accept-reply}
Identically to the propose replies, accept replies are steered to \locentries~(as explained in \S\ref{sec:base:loc}).
The worker periodically polls active \locentries~and takes action when the \locentry~signals that a majority (or more) of replies has been gathered, or if one reply of type \alreadycom/\loglow~has been received.
With 5 machines we wait for 2 replies, since we already have a reply 
-- always an \ack, because we must have accepted locally--
from the local machine. Then the worker acts on the replies as follows. (Note that the list below also denotes the priority with which replies are handled).

\squishlist

\item \alreadycom: <identical to propose> 

\item \loglow: <identical to propose> 

\item \ack: if a majority of \ack~replies has been gathered and none of the above is received, then the \locentry~transitions to \bcast~state, so that it broadcasts commits the next time it is inspected.

\item \highprop/\highacc: <identical to propose>.

\item \loghigh: If a \loghigh~reply has been received and none of the above has occurred then we retry the RMW, transitioning the \locentry~to \retry-state. Retrying is discussed at length in Section~\ref{sec:how:retry}. This is different than proposes, as repeated \loghigh~replies do not lead to a commit. This is discussed at length in \ref{sec:loghigh}.

\squishend

\custvspace\beginbsec{Helping}
If the accept was helping (the \locentry~has a \helpflag~to let us know), then if 
any one nack is received (\ie one of \alreadycom, \loglow, \highprop, \highacc, \loghigh), then we stop helping (lowering the \helpflag) and transition the \locentry~to the \need~state, where it will attempt to \qt{grab} the \kv~to perform its own RMW.

Otherwise, if a majority of \acks~are gathered, we broadcast a commit message by transitioning the \locentry~to \bcasthelp~state (so that it knows to broadcast the value getting helped).

\custvspace\beginbsec{Explanation of priority}
Firstly, if the RMW has already been committed or the \logno~has been used, then there is no reason into inspecting other replies.
Again as before there is a question of priority.
The Paxos invariant is that if a majority of machines ack an accept, then that command is committed.
Therefore, if a majority of \acks~is received, we can prioritize it.  
Otherwise if any of \highprop/\highacc/\loghigh~has been received, the worker will create a TS higher than all received and switch the \locentry~to the state \retry.

Note that it is possible that a majority of machines have acked our accept, but we don't see that (i.e. because we only inspected 2 out of 4 replies and one of the two was a \highprop~reply). One of two things can happen in this case: we will get helped by a remote machine, or we will end-up \qt{helping ourselves} (help is discussed in \S\ref{sec:help}).

\subsection{Commits}
On inspecting the \locentry~and finding it in state \bcast~or \bcasthelp, commit messages are broadcast. 

\custvspace\beginbsec{Why commits}
In principle, an RMW is committed if it has been accepted by a majority of machines.
Still commit messages are necessary. The reason is that it is impossible to know whether an RMW that is accepted, is also committed, unless you help it yourself.
Therefore, without commits before beginning an RMW you would have to first help any previously accepted RMW, even if that has happened a long time ago.
In addition, commits ensure that an RMW can not be committed twice (\eg helped in \lognoeq 5, and then committed by the originator again in \lognoeq 7, is possible without commit messages). Finally, they allow us to reply to the client with the guarantee, that if they read they will see the committed value, without having to run Paxos. In Section~\ref{sec:cor}, we show how commits help us ensure the correctness invariants for moving across \lognos.

\custvspace
Commits include all of the RMW details (its \rmw, \logno~and \val). This allows us to ensure that even machines that have not seen the previous accept messages can commit the RMW.
However machines that have acked this RMW already have this data. We discuss the implementation of an optimization that leverage this observation in Section~\ref{sec:how:commits}.

After broadcasting commits, \locentry~is transitioned to the \committed~state. 
Commits are always acked; upon gathering a majority of acks, we commit the RMW locally and reply to the client that the RMW is completed including the value to be read.
Receiving a commit message always results in an unconditional commit of the RMW.
When committing, we must register the new RMW, if not already registered, we must update the \val, \comlogno~and~\comrmw~fields of the \kv~if no later \logno~has been committed and we must transition the \kv~in \invalid~state, if it currently works on the \logno~to be committed.

\subsection{A Note on the accepted-TS}
You may be tempted to optimize away the \accts~of the \kv, after all it is only used once. Do not. It is an integral part of the Paxos algorithm and critical for correctness. Without the \accts~the proposers have no idea what they should help, and the receivers of proposes do not know what to tell proposers they should help.

In a nutshell, the \propts~is there to block lower proposes and accepts. The \accts~is there to tell proposes what they should help.

\section{The back-off mechanism} \label{sec:backoff}
We saw that if an RMW does not find the \kv~in \invalid~state, it does not grab it. Rather its \locentry~transitions to the \need~state. The \locentry~also stores the RMW that is currently holding the \kv~and its state. 

When the worker inspects the \locentry, if it finds it in \need~state, it attempts to grab the \kv. The \kv~can be grabbed if it's in \invalid~state. 
When the \kv~cannot be grabbed, the \locentry~notes the details of the \kv, its state, its current \propts~etc. Therefore, when the RMW fails repeatedly to grab the \kv, it takes note of whether the RMW that holds the \kv~has made any progress.
If nothing has changed since the the last time that the RMW attempted to grab the \kv, then a counter that is stored in the \locentry~is incremented. The counter measures the time in which the \kv~has not changed. When the counter reaches a threshold (pre-determined \eg at compile-time), then the RMW will attempt to steal or help the entry.
Otherwise, if there is progress since last inspected, then the counter goes to zero.

Essentially, this is a back-off mechanism. Recall that we can have thousands of RMW running concurrently. If a few of them are on the same \kv, then there is no chance of progress.
Instead, we limit the conflicting RMWs to one per machine, in the worst case. This is because a \kv~can be grabbed by a remote machine.

\custvspace

If the counter reaches the threshold, that typically means that the \kv~is held by a remote machine, which has failed. In this instance, if the \kv~is in \proped~state, an RMW can simply \qt{steal} the \kv, by using a higher \propts. The RMW will trigger propose broadcasts, and will transition its \locentry~to \proped~state, whereby it will wait for propose replies.

However, if the \kv~is in Accepted state, we can \emph{not} steal it. We must help it.
\section{Help} \label{sec:help}

If a propose finds a \kv~in \acced~state with a lower \accts, then help must be triggered. The reason is that, it is possible that the propose reached some machines before the Accept. Therefore, the Accept is blocking the propose in some machines, and the Propose is also blocking the Accept in some machines. The proposer cannot know for sure, because it only inspects a majority of responses and therefore it pessimistically breaks the deadlock, by helping the Accept.

A propose helps the accept, by broadcasting accepts with its own TS. This is crucial. The proposer will use its own TS, which is the big one. Why does then a \lowacc~reply has the \accts~inside it, if it's not going to be used? Because the propose must help the highest \accts. Again this is a matter of correctness.

Help can occur when a \lowacc~reply is received to a propose, or when the \locentry~times out on waiting to grab the \kv~locally, and the \kv~is in \acced~state.

\custvspace
\beginbsec{Helping after a \lowacc~reply}
Assume we broadcast a propose for an RMW -- let's call it l-RMW -- and we receive a reply \lowacc~from a different RMW -- let's call it h-RMW. We then attempt to help h-RMW. However, after we have finished helping h-RMW, we must go back to executing our own l-RMW.
To help us, the \locentry~has the following fields: a flag called \helpflag, that is raised to denote that the \locentry~is currently helping h-RMW, and a \helploc, a data structure similar to the \locentry, which will be used to store information about the h-RMW, ensuring that no information about l-RMW is overwritten.

From that point we move as before, attempting to accept h-RMW locally and then commit it.
However, if we are not able to accept locally, or if we receive a \highprop~or \highacc~reply, then the help is cancelled: we take down the \helpflag~and we transition the \locentry~to \need~state, so that it starts over.

If we manage to receive a majority of acks for h-RMW, that means it is committed. We then transition the \locentry~to \bcasthelp~state, denoting that commits must be broadcast, but they must use the \helploc~for the correct value. Finally, when a majority of commit acks have been gathered, we commit to the local KVS, the \helpflag~goes down and the \locentry~transitions to the \need~state. Notably, if we were helping ourselves, we should detect it and simply free the session.

\custvspace
\beginbsec{Help after waiting locally}
In the case, where we timeout on attempting to grab a \kv~locally, which is in \acced~state, the \kv~cannot be stolen (like we would do if it were in \proped~state).
The reason is that the \acced~RMW --let's called it h-RMW -- may be accepted by 3 out of 5 machines and thus committed. It is possible that the issuer had broadcast commits and then died, and thus it is possible that out of the 4 remaining machines, one received the commit and has been reading the RMW, two machines have no idea about the h-RMW and our local machine has \acced~but not received the commit. Therefore, if we were to simply steal the \kv, and put it in \proped~state, we would open the door for other RMWs to be committed in the \logno, where an RMW is already committed.

Therefore, we must act as if we had sent a propose message to the local KVS and it has responded with a \lowacc~reply. The \kv~remains in \acced~state, but its \propts~is advanced to the TS of our propose (for the propose we use a bigger TS than the \kv's \propts). We then broadcast proposes to remote machines.
From that point we proceed as we would normally. Specifically, if we can gather a majority of acks from the remote machines, then we can get the local \kv~\acced~but for our RMW. Otherwise, if we must help, then we proceed to help as described above.

\beginbsec{Crucial Take-away}
It is crucial to remember that a \kv~can never transition from \acced~to \proped~(and remain in the same \logno). It can only transition from \acced~to \acced~with a higher \accts.
\section{Correctness}\label{sec:cor}
We have added several extensions to Paxos to make it run repeatedly (\ie across \lognos) for a given key.

It is challenging to argue that the protocol is both  correct and  efficient.
We will attempt to do this in the following manner. 
First we will list a few invariants. We will argue (or informally prove) that the protocol  enforces said invariants. We will show how additions to the protocol help in doing that.
Then, we will link the invariants with high-level correctness requirements: an RMW commits exactly-once, the issuer of an RMW reads the correct value. We will try to provide arguments that the invariants are both sufficient and necessary.

\subsection{Invariants} \label{sec:cor:invs}

\squishlist
\item If machine M works (proposes or accepts) on \lognox, then it must be that all previous \lognos~have been committed (\emph{inv-1})
\item If machine M works (proposes or accepts) on \lognox, then it must be that M knows what has been committed in \lognoxmin~(\emph{inv-2})
\item It can never be that an RMW gets accepted locally for \lognox, if it has already been committed in a \lognoeq Y, where Y < X (\emph{inv-3})
\squishend

Roughly, we will enforce inv-1 by ensuring that work on \lognox~can start only after \lognoxmin~has been committed.
Inv-2 and Inv-3 will be enforced by nacking with \loghigh~all proposes and accepts that refer to \lognox~when \lognoxmin~is not known to be committed.

\subsubsection{Inv-1}
Inv-1 mandates that a machine M can never work on \lognox, unless all smaller \lognos~have been committed. 

This is proved trivially by the following statement. For machine M to work on \lognox, it must be that at some point one machine grabbed its \kv~for \lognox, with an \invalid~\kv~and \comlognoxmin. There is no other way for work to start on \lognox. 

\subsubsection{Inv-2} 

Inv-2 mandates that machine M cannot work (i.e. propose or accept) on \lognox, unless it knows what was committed in \lognoxmin.
Note that this is subtly different than inv-1, because it mandates that M itself must know what was committed on the immediately previous \logno.

For an RMW to work on a \lognox, it must be that it has grabbed the \kv. There are two cases on how an RMW can grab a \kv.

\custvspace
\beginbsec{1st case} The RMW grabs the \kv~it found in \invalid~state. The invariant is trivially enforced here: the entry cannot be in \invalid~state, unless it knows that the previous \logno~has been committed.

\custvspace
\beginbsec{2nd case}
The second case is that the RMW attempts to steal or help an entry after the back-off timeout expires. However, in the first place, the \kv~can only transition to \proped~or \acced~state iff the immediately previous \logno~is committed, i.e \logno~=~\comlogno~- 1. 
This is because remote proposes/accepts are nacked with a \loghigh~reply if the previous \logno~than the one they want has not been committed. Same goes for local proposes/accepts.

\subsubsection{Inv-3}

Inv-3 mandates that it can never be that an RMW gets accepted locally for \lognox, if it has already been committed in a \lognoeq Y, where Y < X (\emph{inv-3}).
\custvspace

\beginbsec{Proof}
Assume machine M proposes in \lognoeq Z, an RMW it previously locally accepted in \lognox. 
Also assume that the RMW get committed in \lognox, because some other machine helps it.
It suffices to show that M will receive at least one \alreadycom~reply to its propose message.

Firstly, it must be that X < Z - 1, because M, will only attempt to propose on \lognoeq Z if it knows what has been committed in \lognoeq Z -1 (from inv-2), and therefore it has registered the RMW committed in \lognoeq Z-1. 
Furthermore, we know that the RMW has already been committed -- by help --in \lognox, before M proposes for \lognoeq Z. This is because the \lognoeq Z - 1, can only be committed if the \lognox~is already committed (from inv-1).

Therefore, before M can issue proposes for \lognoeq Z, it must be that a majority of machines have already acked an accept for every \logno~<= Z-1.  
Because machines only ack accepts, if they have committed the previous log-no (or else the respond with a \loghigh), we can infer that for each RMW committed in \lognos~smaller than Z, there is a majority of machines that have committed the RMW. Therefore, when M issues proposes for for \lognoeq Z, a majority of machines must have committed and registered the same RMW in \lognox~and thus M must received at least one \alreadycom~reply.

\subsection{Guarantees}
Let's now see how enforcing the above invariants results into higher-level guarantees.
Inv-1 is just there to simplify proving inv-2 and inv-3.
Firstly we will discuss why each of inv-2 and inv-3 are necessary. Then we will see a counter-example of what can happen without them.

\subsubsection{Necessity of inv-2}

Assume we ack accepts, without knowing the previously committed RMW. For example, in a 5-machine deployment, assume machines M1, M2, M3, M4 ack an accept from M5 for \lognoeq 10, but they all have a \comlogno~= 2. This can happen if M5 has committed all the intermediate \lognos. Then assume that M5 dies.
M1 steals the \kv~to help in \lognoeq 10. But M1 cannot work on \lognoeq 10, because it does not know what has been committed in \lognoeq 9, and therefore, if it fails to help, it will have to do its own RMW, but it wont know what value to use as the previously committed.
Therefore, M1 has to start from the \comlogno + 1 (\ie 3) and work its way up.
However, if it sends a propose for \lognoeq 3, everyone else will answer with a \loglow~nack, but they will only include the RMW committed on \lognoeq 2!

\custvspace
\beginbsec{Alternative solution}
It is possible, in the above example to work directly on \lognoeq 9, which we know has been committed (from inv-1) and therefore if M1 issues proposes for \lognoeq 9, it should be able to track down the highest accepted-TS, so that it can commit it directly without bothering with accepts. However, the other machines will still respond with \loglow~nacks to a propose for \lognoeq 9 (because they have accepted M5's RMW to \lognoeq 10). So we then should have to take a lot of extra care (a lot of added metadata and complexity) to have the machines remember the last accepted value/ts for both the current \logno~ and the previous one, so that they can answer to proposes for both \lognoeq 10 and 9.

Note this alternative would also violate inv-3, as no machine will know what has been committed in \lognos~3 through 8. So it is possible that it is one of their own RMWs, and they can end up repeating it, breaking exactly-once semantics.
We could solve this, by having accept messages including a \regedrmw~field that would allow a majority of machines to register committed rmws, even if it has not actually committed them itself. However, that would incur a fixed overhead in all accept messages.

Overall this alternative seems very complicated and with potentially high overhead.

\subsubsection{Necessity of inv-3}\label{sec:cor:inv3-nec}

\custvspace
\beginbsec{Exactly-once semantics}
Inv-3 enforces exactly-once semantics as it ensures that an RMW can never get committed twice in two different logs, by ensuring that it cannot even get accepted.

\custvspace
\beginbsec{The RMW creates correct value}
The RMW creates its value when it gets accepted locally.
For the RMW to create the correct value, it must be that it knows the value committed in the previous log-no. Inv-2 guarantees exactly that.

Note that the additional step, we take to enforce inv-2 was to force the RMW to work on \kv.\comlogno\ instead of on \kv.\logno, when stealing/helping a stuck RMW after a back-off timeout. If we didn't do that then, it would be possible for an RMW to get accepted locally, without knowing the value committed in the previous log. Then, creating an \accval\ based on the current value of \kv~(\ie \kv.value) would be wrong.

\custvspace
\beginbsec{Ensuring that the RMW reads the correct value}
An RMW that gets retried, may be informed that it has already been committed. 
However, remote machines cannot be reasonably expected to say in which \logno\ the RMW has been committed or what value is it supposed to read. To do that, we would require unbounded storage.
Remote machines simply check their registered \rmws\ (bounded storage) and reply back simply that the RMW has been committed.

What value is then a RMW to read, when it learns that it has been committed?
To solve this we leverage a critical insight: for an RMW to be helped it must be that it got locally accepted. This is because, only accepted RMWs can be helped, and an RMW must get accepted locally before it can be accepted by another machine.

We can then combine the insight with the inv-3 (\ie it can never be that an RMW gets accepted locally for \lognox, if it has already been committed in a \lognoeq Y, where Y < X).
Assuming that inv-2 is enforced, then it must be that an RMW can always return the \accval\ from its \locentry, which gets calculated every time the RMW is locally accepted. Plainly, if the RMW got locally accepted in \lognox, if it later finds out that it has been committed, then it must be that it has been committed in \lognox, and thus it can read the \locentry's \accval.

\subsubsection{Counter-example for the invariants}

Let's look at an example of what bad can happen without the invariants. Assume 5 machines: M1, M2, M3, M4 and M5:
\squishenum
\item M1 accepts RMW-1 locally in \lognoeq 1 and then fails to get accept majority
\item But, M2 helps M1's RMW-1, with an accept from majority of M2, M3, M4.
\item M2 broadcasts commits for RMW-1
\item M3 sees the RMW-1 commit and immediately proposes, accepts and commits RMW-3 in \lognoeq 2, with propose and accept majorities of M1, M3, M4
\item M1 sees the committed RMW-3 and then goes on to the next \lognoeq 3 to retry its RMW-1
\item M1's propose  for RMW-1 in \lognoeq 3 gets a majority of acks from M1, M4 and M5
\item M1 accepts locally RMW-1 for \lognoeq 3
\squishenumend

\custvspace \noindent
From this point on M1 will either find out that its \RMW\ has already been committed, in which case it will not know what value to read (even if it could remember previous accepted-values in its local-entry, it wouldn't know which one got helped!) or M1 will manage to commit RMW-1 in \lognoeq 3, breaking the exactly-once semantics.

The problem here is created because M4 acks M1's propose. M4 has not received the commit yet from M2 which helped RMW-1, and thus has not registered RMW-1, yet.
With the \loghigh~nacks to proposes/accepts, M4 can not ack M3's attempt to commit on \lognoeq 2, before it commits and registers the RMW on \lognoeq 1.

\section{Why and how} \label{sec:why}

In this section we are tying loose ends, discussing the why and the how of several implementation choices an optimizations.

\subsection{Why: \alreadycom} \label{sec:why:alreadycom}
Assume M1 sends a a propose/accept for an RMW that has already been committed; M2 replies
back with an \alreadycom~message that contains the entire last committed rmw.

Firstly the \alreadycom~reply is absolutely essential, to ensure exactly-once semantics, seeing as RMWs can be helped. 

When M1 receives the \alreadycom~rep it attempts to commit its RMW locally, using the last-accepted-value and the accepted-log-no from its \locentry.

In addition, M1 needs to ensure that, before it reports completion of the RMW to the client, a majority of machines have committed it.
To do that it may need to broadcast commits from the RMW.
However, if M2 has already committed a later \lognox, than the one M1 is shooting for (\eg \lognoeq Z), that means that the RMW has been already committed in a \logno~<= Z, and thus it is guaranteed to have been committed in a majority of machines (inv-1 in \S\ref{sec:cor:invs}).
M2 includes that information in the opcode, to help M1 avoid unnecessary commits.
Therefore, two different opcodes are used for \alreadycom~rep, as a performance optimization.

\custvspace\beginbsec{Optimization}
We are implementing one more optimization.
If the \acclogno of the \locentry~is  X and its \lognoeq Z, and if Z > X, this means that the RMW has been committed (from a helper) in an older \logno. 
There is then the potential problem, that the RMW has now grabbed the \kv~for the fresh \lognoeq Z, which it did not released when it committed itself on the older \lognoeq X.
That may delay other local RMWs that are attempting to grab the \kv, but are waiting.
To alleviate this, if we detect this case (\locentry.\acclogno < \locentry.\logno, after receiving an \alreadycom reply), we inspect the \kv~and if it's still held in \proped~state we revert it to \invalid~state. (The \kv~cannot be in \acced~for that RMW in this case, because that would mean that an RMW was accepted locally, but then learnt it was committed in a previous \logno, which is impossible -- from inv-3 \S~\ref{sec:cor:invs}.)

Note something very important: this optimization means that a \kv~can transition to \invalid~state without advancing its \comlogno.

\custvspace
It's worth noting, that optimizations are not really important for \alreadycom~replies as they are  triggered only once every 5k to 50k committed RMWs. 

\subsection{Why: Log-too-low} \label{sec:why:loglow}

The \loglow-means that the RMW was attempting to commit in a \logno~that has been used by a different RMW. Therefore, the RMW must go and work in a bigger \logno. As a result the TSes used so far are meaningless because they refer to a specific \logno. The RMW must start fresh. Also it must grab the \kv~from scratch, because the \logno~it used to have is gone as we committed the RMW included in the \loglow~reply.

\subsection{Why: Seen-lower-acc} \label{sec:why:lowacc}

Firstly, why is it okay to handle \lowacc~replies with higher priority than \loghigh~ones?
Because, if anyone has accepted a value locally, then it must be that majority of machines already knows the committed RMWs in previous logs.

There is a case where a \lowacc~reply can be turned into an \ack~purely as a performance optimization.
Specifically, on receiving a propose, if the same \rmw~is in \acced~state, with a lower \propts~and a lower \accts, then we simply return an \ack~to the proposer. This is correct because returning a \lowacc~reply and the \ack~tell the proposer the same exact thing: broadcast accepts using your TS for the RMW you are doing.

\subsection{How: Retrying} \label{sec:how:retry}

An RMW may attempt to retry. This happens when receiving a \highprop/\highacc~to a propose/accept or when receiving a \loghigh~reply.

Here are the steps to retry:
First we check if our RMW has already been committed (\ie registered), (because it may have gotten help). In that case, we need to broadcast commits, to ensure that we will not respond to the client, unless we know that the RMW has been committed by a majority.

Beyond that, there are two cases where the RMW can retry:
1) the \kv~is \qt{still-proposed} \ie it's is in \proped~state for the RMW that is attempting to retry or 2) the \kv~is \qt{still-accepted} \ie it's is in \acced~state for the RMW that is attempting to retry or, 3) the \kv is in \invalid~state, but in a greater \logno.

If the \kv~is still-proposed, or \invalid, then we simply grab the \kv~transitioning to \proped~state(if it's \invalid), then we transition the \locentry~to \proped~state and broadcast proposes. The propose uses a \qt{higher TS} if it has been triggered by a \highprop/\highacc~reply.

\custvspace\beginbsec{Helping myself}
The case where \kv~is still-accepted is more subtle. The \locentry~still transitions to \proped~state and broadcasts proposes, but the \kv~remains in \acced~state. 
The \locentry~treats this as \lowacc~reply, but transitions its \helpflag~to \emph{Propose-locally-accepted} state. 

As with any \lowacc~reply, the \locentry~records the \accts~of the local \kv~and compares it with any other \accts~of any incoming \lowacc~reply.
If no \lowacc~reply, with a higher \accts~is received, and when inspecting the propose replies, the \lowacc~reply is triggered (\ie we have not received a \alreadycom/\loglow or a majority of \acks) then the \locentry, realizes it must \qt{help itself} and therefore it acts as if it had received a majority of \acks. It updates the \accts~in the \kv~and it broadcasts accepts for its RMW.

Notably, if a \lowacc~reply with a higher \accts~is received then the \helpflag~is transitioned back to default (which is \emph{Not-helping}).

\subsection{How: Accepting locally} \label{sec:how:accept}

First let's assume the case where the RMW is not helping, \ie it's trying to get the \kv~to accept itself, not some other RMW.

\custvspace
\beginbsec{Not helping}
We first check the registered RMWs, if the RMW is already committed then the \locentry~transitions to \bcast~state, so that commits will be broadcast using the \accval~and \acclogno~of the \locentry.

The \kv~will accept locally (\ie transition to \acced~state) if it is still in \acced~or \proped~state for this RMW, \ie its \rmw~matches, its \logno~has not moved and its \propts~is the same as that of the \locentry. If that happens, we then decide the new value created by the RMW, and we store it both in the \locentry~and in the \kv~(as \accval).

If the \kv~does not accept locally, it can be because another propose/accept has been received for a different RMW (or even for the same RMW, if someone else is helping us) with a higher TS.
Or it can be because, the \logno~has been committed. In any case, the \locentry~will transition to \need~state, so that it attempts to grab the \kv~the next time it is inspected.

\custvspace
\beginbsec{Helping}
Assume that l-RMW helps h-RMW.
We are trying to locally accept h-RMW, because we have received a \lowacc~reply to a propose message for l-RMW. Note that the \lowacc~reply may have originated from the local \kv~in the case where h-RMW is already locally accepted, but was stuck and the backoff counter timed out, so we issued proposes for l-RMW.

We don't need to check the registered RMWs here -- it's not wrong to do so-- but if h-RMW has been committed, it has to be in the present \logno~and we will not be able to accept locally anyway. We will not need to broadcast commits for h-RMW, if we learn it already has been committed. In all cases where the local accept of h-RMW fails, we will simply stop helping and revert to working on l-RMW, by transitioning the \locentry~to \need~state and lowering it \helpflag.

There are four cases where we should accept h-RMW locally.
1) The \kv~is still in \proped~state for our RMW and exactly as we left it,
2) The \kv~is in \invalid~state but without having advanced its \comlogno~(which is extremely rare but possible, as explained in \S~\ref{sec:why:alreadycom}), 
3) The backoff counter has timeout for an accepted RMW and the \kv~is still in \acced~state, for that same RMW,
4) Finally, it is possible that l-RMW got locally accepted, then it retried, then it received a \lowacc~reply for h-RMW from a remote machine. If h-RMW has a higher \accts~than the \accts~that l-RMW was originally locally accepted with, then l-RMW must help h-RMW.

In all four cases, we accept locally: \ie the \kv~transitions to \acced~state, and it updates its \rmw, its \accts and its \accval.
The \locentry~transitions to \acced~state, and broadcasts accept messages that contain the value and \rmw~of h-RMW, but the TS used by l-RMW (that's just Paxos rules of helping).

If none of the four cases occur, we immediately stop helping and we transition the \locentry~to \need~state, such that it will start the l-RMW from scratch. The reason is that if we cannot get h-RMW to be accepted locally, it must mean that the \kv~is grabbed by some other RMW, or the \logno~has been taken, in either case we need not bother with h-RMW anymore.

\subsection{How: Committing} \label{sec:how:commits}

\beginbsec{Optimization}
If we receive all the acks (one from every machine including the local one) for the accept, then we do not include the value in the commit message.
The optimization is triggered always in All-aboard (so almost in all RMWs in that case), but it has no impact on performance, even though it substantially reduces network usage. The reason is that we are bottlenecked by the CPU and not the network.

The reason we only apply the optimization when \emph{all} remote machines have acked, is that otherwise we would have to send different commit messages to different machines (no value to those who have acked the accept!). However, this is not a good idea: the commit gets inserted in a larger coalesced broadcast message with other commits, accepts, writes etc. That message is then sent to allow remote machines. The invariant is that all remote machines receive the same message. If commits would have variable size w.r.t. the receiver, then we would loose the ability to create these large coalesced messages.

We could also avoid sending the \rmw. However, then the \kv~would have to remember the last-accepted-rmw-id, which is not a big issue (we would just add that field to the \kv). We would then register that last-accepted-rmw-id, which would also be fine: 
in case the last-accepted-rmw-id has been updated in the interim, that has to mean that someone else committed the RMW. However, since there was no performance benefit to removing the value from the RMW, we chose to keep the \rmw~to help with debugging.

\subsection{Why: Log-too-high}\label{sec:loghigh}
Here we discuss 1) why a commit message is triggered after receiving repeated \loghigh~replies and 2) how committing an RMW only after receiving acks for its commit message helps.

It is possible that Machine M1 dies while issuing commits.
M2 is the only machine, that gets a commit and thus M2 blissfully issues a propose in the next slot, but there is a chance that the propose will never get acked: if no other machine is trying to do an RMW on it, then the other machines will not learn of the commit. M2 detects this case after receiving a few consecutive \loghigh~replies for the same propose and broadcasts a commit for the previously committed RMW, by reverting the local-entry to \bcasthelp~and filling the help from \kv~with the last committed RMW (its log-no, value and RMW-id are all known as they must be stored on the \kv).

Note that commits are only triggered by proposes that get \loghigh~nacks. This is because an accept that gets a \loghigh~will result in sending propose messages again (\ie the \locentry~will go to \retry~state).

\beginbsec{Pathological case} 
It can be that we end up in the following pathological case: a propose gets acked but the accepts gets \loghigh, resulting in retrying proposes which get acked and so on for ever. This is very unlikely, because it would mean that repeatedly proposes get acked by a certain majority and accepts by a different one.
One could handle this case be triggering commits of the \comlogno. We have not. 

\beginbsec{Committing after commit acks}
To mitigate the performance impact, we must ensure that machines do not try to propose too fast (we focus on proposes as they come before accepts and are thus more likely to be \qt{early}).
If machines propose timely, after commits have been delivered, then these replies will not trigger in most cases  and therefore no harm, no foul.

When Machine M1 broadcasts commits for \lognox, this allows M2 to receive and apply the commit and then propose on \lognox~+ 1. However, in the common case, there is enough time for the rest of the machines to have received the commit before they receive M2's propose. Indeed, this is the case in our setup.
However, other local threads/sessions within M1, can see the commit much earlier and waste resources on issuing proposes that will receive a \loghigh~reply. 
For this reason, M1 applies the commit to its local \kv, only after it has received one commit ack (we actually do it after a majority of acks, but 1 ack will not be any different).

As a result, there is no performance penalty from adding the \loghigh~replies. Roughly, a single \loghigh~reply is sent once for every 3 thousand committed RMWs.

\section{All-Aboard} \label{sec:all-aboard}

Here we will describe an optimization to Classic Paxos, called All-aboard. All-aboard was proposed in Howard's thesis~\cite{Howard:2019}. We will give a brief description of it, focusing on how to implement it over our existing CP specification.
In a nutshell the  All-aboard optimization does the following:
instead of broadcasting proposes, you can propose locally and then accept locally and broadcast accepts. There are two catches: 1) the accept is successful only if acked by all 2) the accept must have a lower TS than any propose that gets broadcast for the same key and \logno.

We will attempt this only in the first time an RMW runs; if the RMW is not successful in any way (\ie it fails to grab the \kv, or it fails to accept locally, or it fails to gather all the acks), then it simply executes Paxos instead. 99.7\% of completed RMWs are All-aboard, so there is really no reason to try adding it in more places.

When All-aboard triggers, commits are also thinner, as they do not include the value.
Despite removing propose messages and cutting most of the payload of commits the per-machine throughput only improves roughly from 5.5 million RMWs per second to 7.5. This is because the CPU is the bottleneck.

\subsection{All-Aboard Theory}
The theory of All-Aboard is presented in Howard's thesis. We will summarize here.
The theory is underpinned by a rule called Flexible Paxos and its refinement.

\custvspace\beginbsec{Flexible Paxos}
It is only required for propose quorums to intersect with accept quorums.

\custvspace\beginbsec{Refinement} 
Actually, it suffices for proposes to intersect \emph{only} with accepts that have smaller TSes.

\custvspace
All-Aboard takes advantage. It sets up a threshold TS.version. Proposes with a lower TS than the threshold are seen only locally, while accepts must gather all acks. Proposes and accepts with a higher TS than the threshold are seen by a majority.

There are two cases when reasoning a propose must always intersect with accepts with a lower TS: 1) a propose with a TS lower than the threshold will always intersect with \qt{lower accepts}, because \qt{lower accepts} are guaranteed to receive all acks.
2) a propose with a TS higher than the threshold will always intersect with \qt{lower accepts}, because the propose must be acked by a majority and thus overlaps with all accepts.

This allows us to optimistically start the Paxos command using a low TS (lower than the threshold) and avoid broadcasting proposes, going instead directly to accepts. If we are not able to gather all ack for the accept, we can then broadcast proposes using a higher TS (higher than the threshold).

\custvspace
\beginbsec{An aside: Singleton Paxos}
Howard's thesis refers also to the dual of this: gathering all acks for proposes and only accepting locally.
This does not actually work when we are not able to gather all acks for a propose. 
Here's why.
If we gather all acks for proposes for TS = X (similarly to all-aboard), then any propose with TS > X, must gather all acks, otherwise it will not intersect with accepts of TS = X.
Therefore, we cannot do the trick where we start from low TSes, and then revert to bigger ones.
Can we then do the opposite? Start from high timestamps and then if the propose fails revert to lower ones? No, because once nodes see a propose, they cannot ack any propose/accept with lower TS (this is often referred to as \qt{promise}).

Therefore, for the Singleton algorithm we do not have a way to revert to Classic Paxos, and thus it is not really useful. %

\subsection{Specification}
Upon grabbing the \kv~for an RMW for the very first time we transition the \kv~and the \locentry~to \acced~state, and we perform the actions necessary to accept locally: we calculate the result of the RMW and we update the respective fields of the \kv~and the \locentry~(such as \accval, \accts, \rmw~etc.). We also raise a flag \allab~inside the \locentry. 

Since the \locentry~is now in \acced~state we inspect its replies periodically as before.
We know the \locentry~can only move forward if it receives \acks~from \emph{all} machines, because its \allab~flag is raised. 
We handle accept replies as follows.

If we find any nack, we transition the \locentry~according to the guide in Section~\ref{sec:prot:acc-recv}. This is because we need \emph{all} remote machines to \ack~the accept. Therefore any nack must trigger its effect without waiting for more replies, \eg any of \highprop/\highacc/\loghigh~will transition the \locentry\ to \retry~state. 
If we have received only \acks~but not all of them, then we simply increment an a special counter called \allabtimeout, which is a field of the \locentry.
If \allabtimeout~reaches a predetermined threshold, then we transition to \retry~state. 
In the case where the local \kv~is still locally accepted for the current RMW, then the \qt{helping myself} case (described in Section~\ref{sec:how:retry}), gets triggered.

The \allabtimeout~ensures that if a machine is slow (or failed) we will not indefinitely wait for it. Notably this timeout, can be arbitrarily small without any potential to violate correctness. Avoiding false positives however is advisable for performance.

\custvspace
\beginbsec{All-aboard TS}
When executing all-aboard, we must guarantee that the used TS is smaller than the TS of every other propose. This is fairly straightforward; we always use a TS.version = 2, when running all-aboard. Remember that for CP, we always use a TS.version = 3. If all-aboard is not successful, then it will run CP (broadcasting proposes) and will have to use a TS.version >= 3. 

\custvspace
\beginbsec{Note}
A final note -- obvious in hindsight -- but easy to miss, is that if we already suspect a remote machine to be slow/failed, we should avoid triggering the All-aboard optimization all together.
If we do not, and then a machine is unresponsive for a few seconds, then during that time we will be constantly having to wait for the \allabtimeout~to expire. This would be terrible for performance. It is very easy to avoid this. If we have not recently heard of every other machine, we simply execute Classic Paxos for every RMW.

\section{Adding writes} \label{sec:writes}

In this section, we will discuss the implementation specification of ABD writes using carstamps. %

Writes need not solve consensus; in contrast concurrent writes can execute and be serialized post-hoc deterministicaly in each node using logical timestmaps (\ie TSes). Therefore, writes can fundamentally be implemented more efficiently than RMWs.
We can do roughly 5.5 million Classic Paxos RMW/s per machine (5 machines).
All-aboard does roughly 7.5 million and ABD writes reach 12 million.

Therefore, there is benefit coupling them together.
If the client wants 90\% of the time to do simple writes and 10\% of the time to do RMWs, then we can increase performance, by using ABD 90\% of the time. 

To do so, we use a technique called \emph{carstamps} to be able to serialize Paxos RMWs and ABD Writes.  
Below we dive into how carstamps interact with the Paxos protocol and what are the necessary additions to support them.

\custvspace
\beginbsec{Basic idea}
The \kv~will have a \basets. Writes will increase the \basets. RMWs will choose both a \logno~and a \basets. The  \basets\ (which is Lamport clock same as all TSes we have discussed) along with the \logno\ comprise the carstamp. For example imagine key A and machines M1, M2, M3, M4, M5. M1 writes A increasing its \basets~to {1, M1}. M2 performs an RMW on A choosing \lognoeq 1 and \basets~= {1, M1}. M3 writes A before M2 finishes, increasing the \basets~to {2, M3}. M2's RMW precedes M3's writes and as such will only be applied by machines that have not seen M3's write. Assume that M2's RMW fails because M4's RMW wins out. M4 may choose \basets~={2, M3}, and \lognoeq 1.

\subsection{Invariants}
Firstly, we need to make sure that the issuer of an RMW reads the correct value, and the RMW overwrites the same value in all machines, despite getting helped.
We will achieve that effect in the same spirit as we have done so far.
When the RMW gets accepted locally, then it selects its \basets. If it is to get committed -- help or not -- it must be that it gets committed with that same \basets. This guarantees, that an RMW always gets committed with the same \basets~in all machines.

Secondly, we need to make sure that the RMW overwrites the most recently committed write.
Here is why this is slightly harder.

\custvspace
\beginbsec{Enforcing the Invariants}
When accepting locally, we choose the value that will be read. If the RMW is committed in that same \logno, then that value is set in stone, \ie~helpers cannot change it.
For example, imagine that  Machine-1 accepts locally, then it immediately becomes unresponsive and Machine-2 helps it. When Machine-1 becomes responsive again it realizes its RMW has been helped, but it needs to know what value to read. There are two invariants of Paxos that help us then read the correct value: the RMW can only be helped in a \logno~where it accepted locally and the value that was helped was the one accepted locally. Therefore, it is safe to read the locally accepted value. This is described in detail in Section~\ref{sec:cor}.

\custvspace
Therefore, to ensure the second invariant \ie to maintain linearizability with writes and RMWs, it must be that when accepting locally, the RMW uses a \basets~that is bigger or equal than that of any write that had completed before the RMW was issued.
I.e. it must never be that a write completed before the RMW began executing, but was not seen by the RMW.

To achieve this we include the \basets, in propose messages, asking other machines if they have seen any more writes. When a remote machine intends to ack the propose (\ie none of the nack conditions are triggered), then it also inspects its \basets; if the propose's \basets~is smaller that what locally stored then the propose is still acked, but the \ack~reply contains the value and its \basets. This allows the proposer to ensure that it will use a \basets~that is at least as big as the biggest write that has completed. Sometimes the propose, will find the ts of writes that have not yet completed. This is okay: if the RMW completes it is as if its base write is also completed (similarly to the second round of an ABD write or read).

\subsection{What about All-aboard}

Unfortunately, All-aboard immediately selects the value that it will overwrite, as it gets immediately accepted locally. This however violated the second invariant that denotes that the RMW must overwrite any completed write. The problem is that there may be completed writes that have not been received locally.  The fix would be to add a broadcast round that reads timestamps similar to what ABD-writes do. However, that seems awfully close to having proposes and thus beats the purpose of All-aboard in the first place. The inverse optimization, Singleton Paxos, where proposes are acked by all instead of accepts, would naturally work with carstamps. As explained, in Section~\ref{sec:all-aboard}, Singleton Paxos is not possible to deploy if all acks cannot always be gathered.

We have not added the first round to All-aboard. %

\subsection{Specification Changes}

\custvspace\beginbsec{Metadata changes}
\squishlist
\item The \kv~has a \basets~field, to be used by ABD writes for serialization, but also by RMWs, to serialize with writes.
\item The \kv~has an \accbasets~field, which is used to return to proposes when sending them \lowacc~replies, along with the \accts~the \accval~and the \rmw~(which is the rmw-id of the last accepted RMW). This will allow helpers, to commit an RMW with the correct \basets.
\item \loglow~replies include the \basets~of the last committed RMW. Finally this is also useful when receiving commits without any value (see \S\ref{sec:how:commits}), to know which \basets~to commit.

\item The \locentry~needs a \basets~field that specifies the chosen \basets~at local-accept time.
\squishend

\subsubsection{Protocol changes}

\custvspace\beginbsec{Sending Proposes}
Proposes now include a \basets~field. 

\custvspace\beginbsec{Proposes replies}
On sending propose replies we make the following changes
\squishlist
\item The \lowacc~reply includes the \accbasets~of the \kv, which is the \basets~of the RMW that may be helped.
\item The \loglow~replies also include the \basets~of the \kv. 
\item A new type of ack is introduced: \ackbase~ which is triggered every time  a propose is is received that can be acked but contains a low \basets~compared to the locally stored. The payload of the reply includes a value and a \basets. 
\squishend

On receiving a propose reply of type \ackbase, the proposer overwrites its locally stored value and its \basets, if the \basets~of the \ackbase~is bigger than the locally stored \basets. 

\custvspace\beginbsec{Optimization} As an optimization, we note on the \locentry~(by raising a flag) that the RMW has looked for a fresh \basets~and therefore subsequent proposes from this propose (because it may get retried), need not inspect the \basets~of remote nodes

\custvspace\beginbsec{Accepts} Accept messages include the \basets. This gets written in the \kv's \accbasets~field. There is no change to accept replies.

\custvspace\beginbsec{Commits} Finally, commits include the \basets~of the RMW to be committed.  Recall on Section~\ref{sec:how:commits}, we discussed an optimization where commits can be broadcast without their value, if the accept has seen \acks~from all machines. Those commits need not include \basets~either. The receiver side has stored the \basets~in its \accbasets~field of the \kv~and will use that. A potential pitfall: if the \kv~is not still in \acced~state, but it has progressed, its \accbasets~should not be used.
\section{Adding Reads}\label{sec:reads}

Using Paxos to perform reads is far from ideal. Paxos delivers exclusive access to a key to one session. This is 1) very costly, as it is fundamentally hard to do and 2) it hinders concurrency. Reads can be fully concurrent and need no coordination with each other. To give you a sense, 
we can do about 140 million ABD reads per second, in contrast to 27 million CP RMWs. 

Below, we will explain ABD reads (from this point on simply \qt{reads}) with carstamps.

\subsection{Protocol}
Firstly the reader broadcasts its locally stored carstamp (\logno~plus a \basets).
The receiver inspects the \basets~and \comlogno~of the \kv~and it sends back one of three possible replies:
\begin{enumerate}
    \item \carstamplow: if the incoming carstamp is lower than the locally stored, then the reply includes the locally stored carstamp along with the locally stored value.
    \item \carstampequal: if the incoming carstamp is equal to the locally stored, then the reply has no payload
    \item \carstamphigh: if the incoming carstamp is higher to the locally stored, then the reply has no payload
\end{enumerate}

The reader collects a majority of answers and reads the highest carstamp received.
If the reader is not certain that a majority of machines store the value that is about to be read, then before reading, it  first broadcasts a commit (as in a Paxos commit) including the carstamp and the value that will be read.

\custvspace
\beginbsec{Commits and reads}
Earlier we said that commits are needed so that reads know what is committed.
What would the alternative be? Could reads simply look at the values that are accepted and read them? After all if a value is accepted by a majority then it is in principle also committed. However, the accepting majority and the majority that replies to a read may not be the exact same. In which case reads will be unable to infer whether a value that is accepted in one of the replies, is indeed accepted by a majority. To solve this problem the reads themselves would have to \qt{help} accepted values by broadcasting accepts. That would complicate the read protocol, because as we know the path from sending accepts to getting a value committed can be pretty complex.

We want to avoid having reads running any part of the Paxos protocol and that's where commit messages are so useful. Notably, reads may be forced to broadcast commits, but commits can only get acked by remote nodes.

 \section{Conclusion}

This document described in detail a high-performance implementation of Classic Paxos. We also saw how it can be combined with ABD writes and reads and how it can be optimized to include All-aboard Paxos.

\bibliographystyle{IEEEtranS}
\bibliography{paper}

\end{document}